# Multi-Principal-Element Approach to High-Performance Thermoelectric Materials

S. Joseph Poon[1] and Jian He[2]

[1] Department of Physics, University of Virginia, Charlottesville, Virginia 22904-4714, USA
[2] Department of Physics and Astronomy, Clemson University, Clemson, South Carolina 29634-0978, USA

## ABSTRACT

High-entropy alloys are characterized by high configurational entropy. Since the discovery of high-entropy alloys (HEA) in 2004, entropy engineering has provided a promising direction for exploiting composition, lattice disorder, band structure, and microstructure effects to advance thermoelectric performance. This review discusses the impact of entropy on thermoelectric properties and looks back at the role of multi-principal-element alloys, a weaker version of HEA, on the development of compositionally complex thermoelectric alloys in achieving high thermoelectric performance. The experimental and theoretical efforts in a wide range of material systems such as TAGS, LAST, half-Heusler, liquid-like copper chalcogenides, SnTe, and $CuInTe_2$ chalcopyrites provide insights into the entropy engineering approach and also promise an emerging paradigm of high-entropy thermoelectrics.

Key words: Multi-principal-element alloys, multi-element alloys, composition, entropy, thermoelectric properties, figure of merit, state-of-the-art thermoelectric materials.

Correspondence: sjp9x@virginia.edu, jianhe@g.clemson.edu

## I. INTRODUCTION

More than fifty percent of the usable energy produced worldwide is wasted in the form of heat (Johnson et al., 2008). Thermoelectric conversion is a technique for recovering waste heat as useful energy through direct heat-to-electrical energy conversion and Peltier cooling, while leaving minimal footprint on the environment (Goldsmid, 2010). Critical to the development of efficient thermoelectric (TE) devices are high-performance TE materials that can be utilized in the temperature ranges of the heat sources

The dimensionless figure of merit ZT, an important parameter that determines the energy conversion efficiency of a thermoelectric material, is defined as

$$ZT = (\alpha^2 \rho/\kappa)T \qquad (1)$$

where $\alpha$ is the Seebeck coefficient, $\rho$ the electrical resistivity, and $\kappa$ the thermal conductivity. $\kappa$ is the sum of $\kappa_e$ and $\kappa_L$, the electronic and lattice contribution to $\kappa$, respectively. $\kappa_e$ is given by $\kappa_e = LT/\rho$, where $L$ is the Lorenz number. Central to thermoelectric material research is decoupling the otherwise adversely interdependent parameters $\sigma$, S, and $\kappa$. Doping is the routine to decouple these parameters and optimize the material's thermoelectric performance: dopants introduced in the parent compound simultaneously optimize the carrier concentration and scatter heat-carrying phonons in line with the classic electron-crystal phonon-glass paradigm (Slack, 1995). In a crystalline parent compound, substitutional dopants break the translational and/or rotational lattice symmetry, yielding site occupational disorder. The level of such disorder is quantified approximately by the configurational entropy $\Delta S = k_B ln\Omega = -N_A k_B \Sigma_{i=1}^{n} x_i ln x_i$, where $\Sigma_{i=1}^{n} x_i = 1$, $k_B$ is the Boltzmann constant, $\Omega$ is the number of probable atomic occupations, $n$ is the number of the substituted components, $x_i$ is the mole content of the $i_{th}$ component, and $N_A$ is Avogadro's number.

The common definition of high-entropy alloys (HEA) is that the crystalline solid solution contains at least 5 constituent elements with contents in the range of 5-35 at. % for each, preferably in equimolar ratios, so as to maximize the configurational entropy (Yeh, 2004; Cantor, 2004). These numbers are believed to be where the energy due to configurational entropy exceeds the formation enthalpy of competing intermetallic phases, resulting in a single solid solution phase (Yeh, 2016). However, it has been shown that the outstanding structural properties of HEA could also be achieved in multi-principal-element alloys (MPEA) (Li, 2019), a weaker version of HEA (Senkov, 2015) based on composition. Thermoelectric parent compounds are usually composed of $\zeta$ base components, where $\zeta \geq 2$. By adopting herein a configurational entropy based definition, thermoelectric MPEA are to contain at least $\zeta+2$ components, and in order to distinguish from conventional doped semiconductors and semimetals, each multi-principal component must account for at least 0.05 fraction of the parent compound component or substituted base component. However, exceptions to this composition rule can be made for parent compounds that exhibit liquid-like phonon behavior (Chen, 2018).

MPEA are a new alloy design concept for exploiting the vast compositional space of structural and functional materials. In this article, we will begin with a discussion of the fundamentals of entropy engineering as applied to thermoelectric alloys. Firstly, the extreme lattice distortions due to strong chemical disorder scatter phonons, resulting in the suppression of lattice thermal conductivity (Carruthers, 1959). Furthermore, the high lattice strain in MPEA tends to promote phase separation, creating a microstructure to scatter phonons on the mesoscopic scale (Biswas, 2012). We will then revisit some multi-element thermoelectric alloys prior to the discovery of high-entropy alloys. Finally, we will devote the rest of the article to review the current status of multi-principal-element thermoelectric materials.

## II. ENTROPIES AND THERMOELECTRIC PROPERTIES

As illustrated in Figure 1, entropy in its various guises has a profound impact on all thermoelectric properties. First, the configurational entropy simultaneously governs the carrier mobility in the electrical conduction and the phonon mean free path in the thermal tranport, constituting one of the major trade-offs in thermoelectric materials research [Zhu et al., 2017]. Other forms of entropy are more directly associated with transport properties. Specifically, the Seebeck coefficient is essentially the average entropy transported per unit charge. The relation between the charge flow and the accompanied entropy flow is described by the Lorenz number in the Wiedemann-Franz relation. In addition to the configurational entropy, vibrational entropy is also relevant to thermoelectricity, embodied in the specific heat and temperature [Fultz, 2010]. Anharmonicity, which governs the phonon lifetime, is reflected in the difference between the isobaric specific heat $C_p$ and the isochoric specific heat $C_v$. Note that entropy creation (e.g., the Joule effect) would make a thermoelectric process thermodynamically irreversible. Hence, pursuing higher ZT in a material is no more than reducing the entropy creation in the thermoelectric process therein [Goupil et al., 2011].

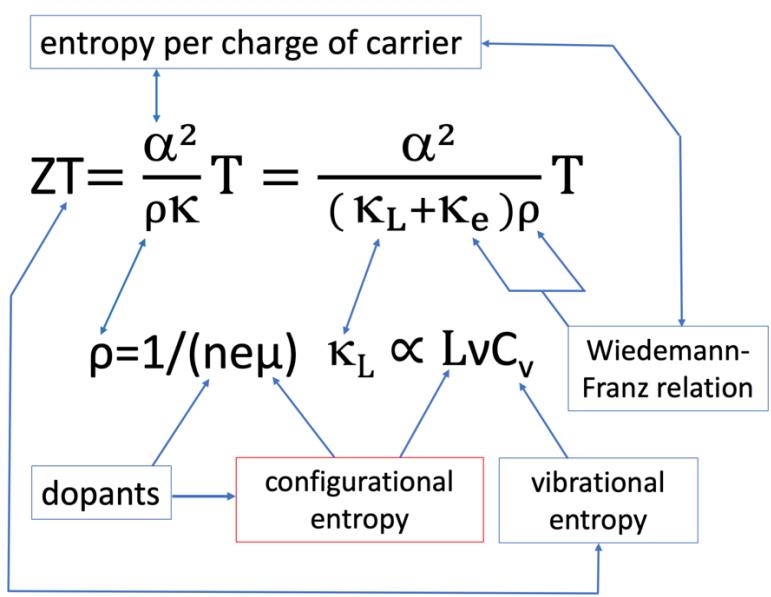

Figure 1. A schematic entropic view of thermoelectric properties. Several important mechanisms such as the electron-phonon coupling and spin entropy are left out for brevity. The n, μ, L, v, and $C_v$ denote the carrier concentration, carrier mobility, phonon mean free path, speed of sound, and isochoric specific heat, respectively.

Table 1 sheds more light on the inherent link between thermoelectrics and MPEA by presenting the four core effects of MPEA and their "translations" in thermoelectrics. Despite no consensus on the scope of validity and applicability of these core effects, especially the entropy stabilization of high symmetry crystal structures and sluggish diffusion kinetics (Pickering and Jones, 2016), it is evident that these core effects are in principle thermoelectrically favored. It is thus justified to treat the configurational entropy as a useful performance indicator in thermoelectric materials research (Liu et al., 2017) and to implement entropy engineering towards a new paradigm of "high-entropy" thermoelectrics. However, a common downside of these seemingly beneficial core effects is that they tend to degrade the carrier mobility, a crucial microscopic transport parameter in the ZT and also in the B-factor that embodies the primary requirements for improving thermoelectric performance. The B-factor is given as follows (Nolas, 2001):

$$B \sim \mu m_d^{*3/2} T^{5/2} / \kappa_L \qquad (2)$$

where $m_d^*$ is the total density of states (DOS) effective mass. The carrier mobility $\mu$ depends on the effective masses, as follows:

$$\mu \sim \frac{1}{m_c^* m_b^{*3/2} \varepsilon^2} \qquad (3)$$

where $m_b^*$ and $m_c^*$ are the DOS effective mass and conductivity effective mass, respectively, and $\varepsilon$ is the acoustic deformation potential for carrier-phonon scattering. The effective masses $m_b^*$ and $m_d^*$ are related by $m_d^* = N_v^{2/3} m_b^*$, where $N_v$ is the valley degeneracy. The B-factor therefore can be written as $B \sim N_v T^{5/2}/(m_c^* \varepsilon^2 \kappa_L)$. Increasing the deformation potential due to lattice deformation and defects reduces both $\mu$ and $\kappa_L$. Thus, there is apparently a trade-off between the lattice thermal conductivity and the mobility of charge carrier in MPEA. So, *how high the configurational entropy ought to be in thermoelectric materials*? Here we argue that the configurational entropy should be high enough to elicit one or more entropy-enabled core effects of HEA while low enough to retain a reasonable carrier mobility.

Table 1. Four entropy-enabled core effects in HEA [Pickering and Jones, 2016] are largely aligned with the material requirements for better thermoelectrics.

| Entropy-enabled core effects in MPEA | The translations in thermoelectrics |
|---|---|
| *(1) Entropy stabilizes solid solutions with high lattice symmetry and leads to extended solubility* | High band degeneracy and extended doping range |
| *(2) Sluggish diffusion kinetics* | Rich multi-scale microstructures |
| *(3) Severe lattice distortion* | Reduced lattice thermal conductivity and degraded carrier mobility |
| *(4) Cocktail effect* | Synergistic effect |

### III. MPEA APPROACH TO THERMOELECTRICS

A recent review on high-entropy functional materials has provided some initial results on thermoelectric MPEA (Gao et al., 2018). MPEA approach sets a new paradigm for exploiting composition effect, and thus high entropy effect, on thermoelectric properties in designing high-performance thermoelectric materials. MPEA approach can be utilized at two levels. At one level, the approach can be employed to optimize thermoelectric properties. At another level, it can be unleashed to discover new thermoelectric materials. The effort to discover high-ZT materials lie in the realm of inverse design that poses great challenges, given the enormous compositional space of MPEA. Inverse design is about solving constrained-satisfaction problems (Arroyave et al., 2016) that have no general methods for finding the solutions. To date, thermoelectric materials research has been largely focused on optimization.

## IV. MULTI-ELEMENT THERMOELECTRICS IN RETROSPECT

From historical perspective, the utilization of multi-element-alloy approach to optimize thermoelectric performance is not new. Researchers in the 1960s explored multi-element doping and alloying to enhance ZT. Thus, it is not surprising that some GeTe-based thermoelectric alloys (Rosi et al., 1960; Plachkova, 1984) can be recognized as MPEA according to the composition rule. Besides GeTe, we will also highlight some early notable thermoelectric MPEA. More generally, MPEA typically have lower thermal conductivities that in some cases can result in higher figures of merit.

(i) Te-Sb-Ge-Ag (TAGS) compounds with the rock salt structure are pseudo-binary GeTe-AgSbTe$_2$ alloys (Rosi et al., 1960; Plachkova, 1984). TAGS-85 (85% GeTe + 15% AgSbTe$_2$) alloys were successfully applied for direct conversion of heat energy into electricity by radio isotopic thermoelectric generators in space (Wood, 1988). The ZT of TAGS was increased from ~1 to above 2 by increasing the composition complexity (Perumal et al., 2016). Figure 2 shows the crystal structure of TAGS in which Ag and Sb co-occupy the Ge sites in the Ge-Te lattice (Snykers et al., 1972). The decrease of thermal conductivity with increasing composition complexity is also noted.

(ii) In the early 2000s, half-Heusler Hf$_{0.5}$Zr$_{0.5}$Ni$_{0.8}$Pd$_{0.2}$Sn$_{0.99}$Sb$_{0.01}$ compounds were found to have lower thermal conductivity and higher ZT~1 compared with ternary HfNiSn and ZrNiSn base compounds (Shen et al., 2001). Another MPEA, (Zr$_{0.65}$Hf$_{0.35}$)(Co$_{1-x}$Pt$_x$)(Sb$_{1-y}$Sn$_y$), showed much lower thermal conductivity compared with ternary ZrCoSb (Xia et al., 2000). Figure 3 shows the systematic decrease in thermal conductivity with increase in the number and concentration of alloying elements. The crystal structure of MPE half-Heusler compound is shown in the left side of Figure 3.

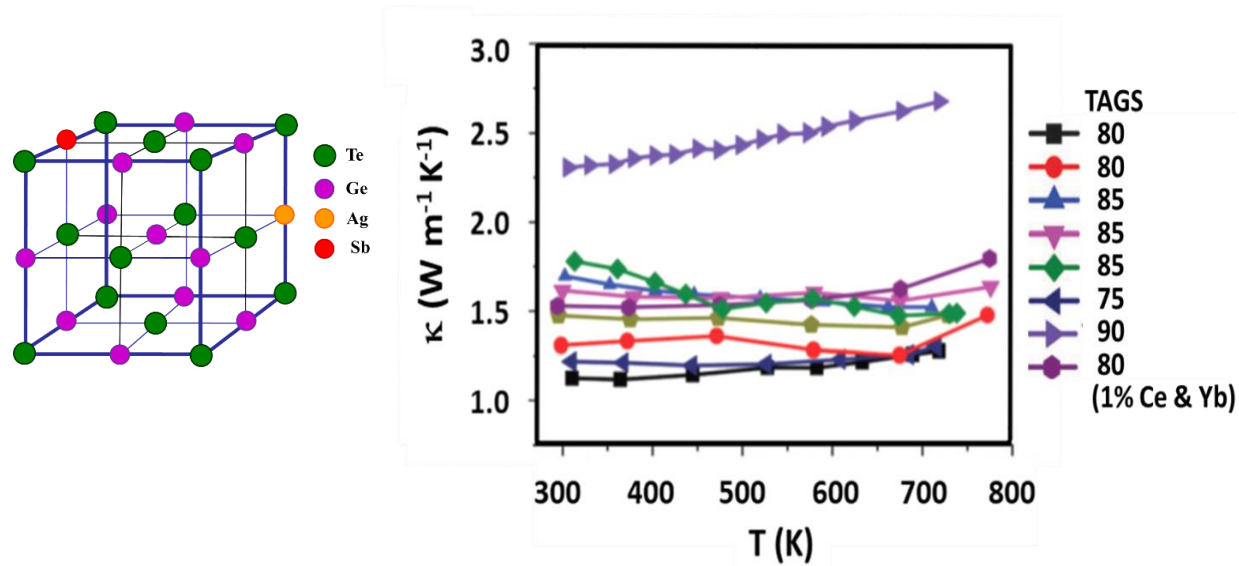

Figure 2. Left: Crystal structure of multi-element TAGS compounds in which Ag and Sb co-occupy the Ge site. Right: Thermal conductivity versus temperature for various TAGS-90, 85, 80, and 85 showing κ decreases with increasing fraction of different alloying phases (Perumal et al., 2016). Right figure is reproduced by permission of The Royal Society of Chemistry https://doi.org/10.1039/C6TC02501C.

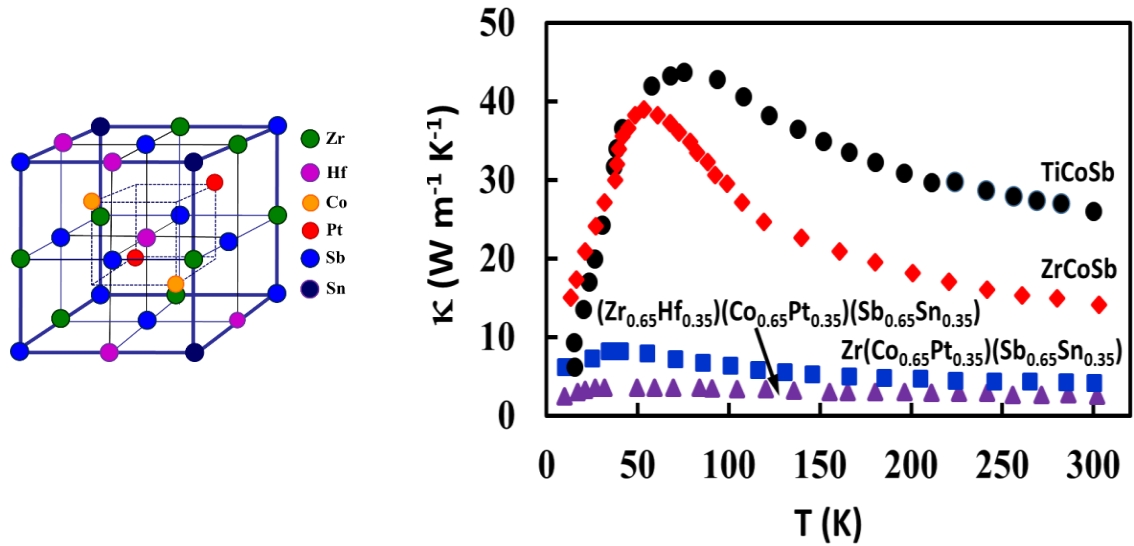

Figure 3. Left: Crystal structure of MPE ZrCoSb-based half-Heusler compounds with the C1b structure. Right: Thermal conductivity versus temperature for four half-Heusler compounds with increasing compositional complexity (top to bottom plots). Right figure is reproduced from Xia et al., 2000. J. Appl. Phys. 88, 1952-1955 with the permission of AIP Publishing.

Besides MPEA, TE alloys that contain two or more dopants investigated in the early 2000s also showed the benefit of entropy engineering. $Pb_mSbAgTe_{m+2}$ (LAST-m) chalcogenides based on PbTe with the rock salt structure were found to have low thermal conductivity below 1 W $m^{-1}$ $K^{-1}$ and high ZT near 2 (Hsu et al., 2004). These alloys were found to be inhomogeneous at the nanoscale with at least two coexisting phases. The minority phase is richer in Ag and Sb, while the majority phase is more deficient in Ag and Sb (Quarez et al., 2005). In another alloy system, $CoSb_3$-based skutterudites co-doped with rare-earth Eu and Ge achieved a high ZT above 1 in $Eu_{0.42}Co_4Sb_{11.37}Ge_{0.50}$, the highest reported for skutterudites in the early 2000s (Lambertson, Jr., et al., 2002). Eu atoms filled the voids inside the Sb polyhedra to act as rattlers. The high ZT was attributed to two doping effects, namely, scattering of low frequency phonons by the low frequency acoustic modes of the Eu rattlers and increase in electrical conductivity due to Ge dopants.

## V. STATE-OF-THE-ART THERMOELECTRIC MPEA

An apparent correlation between ZT and configurational entropy was noted recently (Gao et al., 2018; Liu et al., 2017). High entropy of mixing promotes structural stability, thereby allowing more compositional varieties to be explored. Furthermore, the MPEA principle facilitates elemental diversity, and as such, it enables the control of microscopic parameter and microstructure for obtaining the materials on demand. Figure 4 shows the ZT-T regions of five types of thermoelectric compounds for which MPEA results are reported. These regions are plotted using information from recent review articles (Perumal et al., 2016; Liu et al., 2017; Gao et al., 2018; He and Tritt, 2017; Poon, 2019). More specific sources are cited below. The results show that thermoelectric MPEA have high ZT comparable to that of state-of-the-art materials.

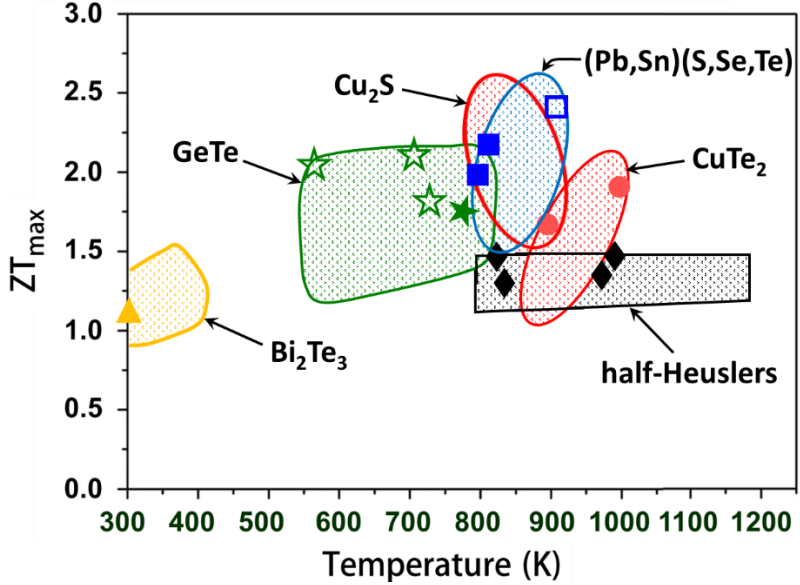

Figure 4. ZT-T regions show maximum ZT ($ZT_{max}$) for five types of thermoelectric compounds: Cu-based chalcogenides in two regions to include $Cu_2S$ type and $CuTe_2$ type, Sn- and Pb-based chalcogenides, GeTe, half-Heuslers, and $Bi_2Te_3$. The regions are identified accordingly as shown. Only ZT near and above 1 are shown. MPEA and multi-element alloys with the highest $ZT_{max}$ are represented by filled symbols and unfilled symbols, respectively: Cu chalcogenides (circle), Sn and Pb chalcogenides (square), GeTe (star), half-Heuslers (diamond), and $Bi_2Te_3$ (triangle). The compositions of these alloys are given below:

The high-ZT MPEA (filled symbols) shown in Figure 4 include: $Bi_{0.5}Sb_{1.5}Te_{2.7}Se_{0.3}$, 1.12, 303 (Xu et al, 2012), where the first number is ZT and second number is the temperature at which $ZT_{max}$ occurs; $(GeTe)_{80}(AgSbTe_2)_{20}$, 1.75, 773 (Davidow and Gelbstein, 2014); $(PbTe)_{0.84}(PbSe)_{0.07}(PbS)_{0.07}Na_{0.02}$, 2, 800 (Korkosz et al., 2014); $Pb_{0.92}Mg_{0.08}Te_{0.8}Se_{0.2}Na_{0.02}$, 2.2, 820 (Fu, 2016); $Cu_{0.8}Ag_{0.2}In_{0.5}Ga_{0.5}Te_2$, 1.6, 900 (Liu et al, 2017); $Cu_{1.95}Ag_{0.05}S_{1/3}Se_{1/3}Te_{1/3}$, 1.9, 1000 (Liu et al, 2017); $Ti_{0.5}Zr_{0.25}Hf_{0.25}NiSn$, 1.5 820 (Rogl et al., 2018); $Hf_{0.65}Zr_{0.25}Ti_{0.15}NiSn_{0.995}Sb_{0.005}$, 1.3, 830 (Chen et al., 2016); $ZrCoBi_{0.65}Sb_{0.15}Sn_{0.20}$, 1.4 970 (Zhu et al., 2018); $Ta_{0.74}V_{0.1}Ti_{0.16}FeSb$. 1.52, 973 (Zhu et al., 2019). Multi-dopant or multi-element alloys (open symbols) include $(Ge_{0.87}Pb_{0.13}Te)(Bi_2Te_3)_{0.05}$, 2.1, 575 (Gelbstein and Davidow, 2014); $(GeTe)_{0.937}(Bi_2Se_{0.2}Te_{2.8})_{0.063}$, 2.2, 725 (Koenig et al., 2015); $(GeTe)_{19}Sb_2Te_3(CoGe)_{0.22}$, 1.85, 730 (Fahrnbauer et al., 2015); $(PbTe)_{0.9}(SrTe)_{0.08}Na_{0.02}$, 2.4, 900 (Tan et al., 2016). Not plotted in Figure 4 due to the lack of MPEA composition are $CoSb_3$ ($CoAs_3$ phase) based skutterudite compounds and magnesium-antimony $Mg_3Sb_2$ ($La_2O_3$ phase) compounds. These materials can achieve high ZT with two or more dopants. Examples include multi-filled skutterudites $(Ba_{0.08-0.125}La_{0.05-0.125})Yb_{0.04-0.25}Co_4Sb_{12-12.5}$ with ZT~1.7-1.9 (Shi et al., 2011; Rogl et al., 2014) and $Mg_3(Sb_{1.48}Bi_{0.48}Te_{0.04})$ with ZT~1.65 (Zhang et al, 2017).

## VI. THE PHYSICS OF THERMOELECTRIC MPEA

In this section, we shall highlight results from recent studies of multi-principal-element thermoelectric alloys. The mechanisms of thermoelectric properties in these compositionally complex alloys are discussed, with focus on delineating the elemental contributions to TE properties wherever possible. The two main groups of thermoelectric MPEA studied to date include half-Heusler compounds and chalcogenide compounds. These MPEA are discussed in the following separate subsections.

(A) Half-Heusler Compounds

Although thermoelectric half-Heusler (HH) alloys are not specifically referred to as MPEA, the core effects of entropy engineering have played important roles in the development of these alloys, much like other compositionally complex materials studied earlier. HH compounds began to attract attention in the late 1990s, nearly a decade after the discovery of large Seebeck coefficient in refractory metals based HH alloys (Aliev et al., 1990). Steady progress was made, which led to ZT~1 in some multi-element alloys (Xie et al., 2013; Chen and Ren, 2013; Poon, 2019). Figure 5 shows the highest ZT for various HH alloys reported since 2000 (He and Tritt, 2017; Poon, 2019). For comparison, HH alloys that contain only one additional alloying element tend to show lower ZT. The progress in ZT can be attributed to improvements in material synthesis and design. Nevertheless, the fact that the majority of HH alloys with high ZT>1 are MPEA underscores the effectiveness of entropy engineering.

To illustrate the multi-principal-element effect, Table 2 shows the results for MPE half-Heusler compounds with the highest ZT (≥1.2) reported to date. The roles of alloying elements in improving the thermoelectric properties are highlighted. Various effects due to chemical disorder and phase separation have clearly increased the scattering of phonons across length scales, resulting in the reduction of thermal conductivity. In addition, band structure design has proven to play a significant role in enhancing the Seebeck coefficient, which resulted in an overall enhancement of ZT. Details can be found in the references cited. While MPE approach has played a major role in achieving high ZT in the HH compounds, comparable result was also obtained in some multi-element and even single-doped compositions (Figure 5), including $Hf_{0.6}Zr_{0.4}NiSn_{0.995}Sb_{0.005}$, 1.2, 860 (Chen et al., 2015); $Hf_{0.594}Zr_{0.396}V_{0.01}NiSn_{0.995}Sb_{0.005}$, 1.3, 900 (Chen et al., 2017); and $Zr_{0.95}M_{0.05}Ni_{1.04}Sn_{0.99}Sb_{0.01}$ (M=$Ti_{0.25}Hf_{0.25}V_{0.25}Nb_{0.25}$), 1.2, 850K (Gong et al., 2019).

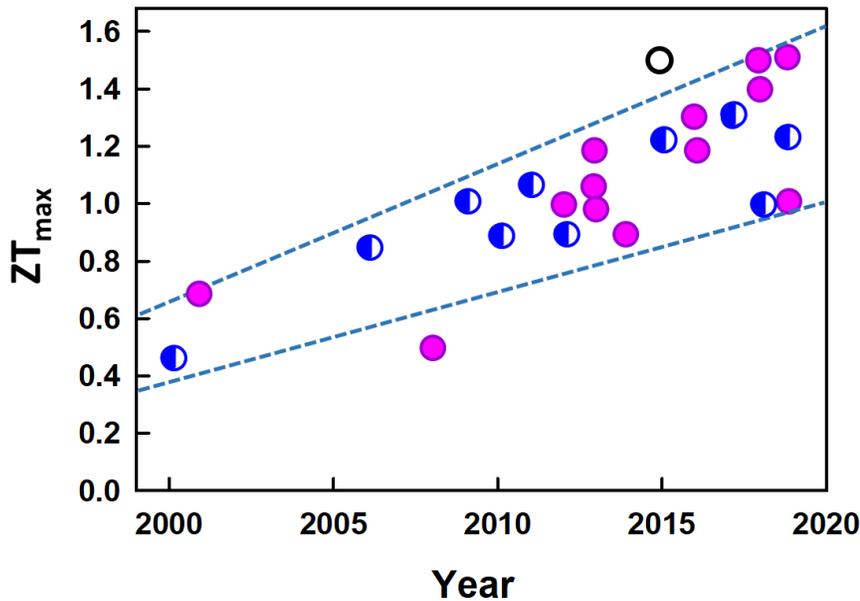

Figure 5. Plot shows advance in the ZT of half-Heusler compounds over time. MPEA are labeled by filled circles. Multi-element alloys are shown in half-filled circles. Open circle represents single-doped composition. The data plotted are accessed from recent review articles and many references cited therein (He and Tritt, 2017; Poon, 2019). Each symbol may represent several alloys with the same ZT.

Table 2. Half-Heusler compounds with high ZT. Enabling mechanism features the roles of alloying elements including those that are dopants.

| Composition | $ZT_{max}$ | Enabling mechanism | Refs. |
|---|---|---|---|
| $Ti_{0.5}Zr_{0.25}Hf_{0.25}NiSn$ n-type | 1.2 | Phase separation due to Hf/Ti atomic size mismatch and mass and strain fluctuations from Ti/Zr/Hf enhance phonon scattering | (Schwall and Balke, 2013) |
| $Hf_{0.65}Zr_{0.25}Ti_{0.15}NiSn_{0.995}Sb_{0.005}$ embedded 2% $ZrO_2$ n-type | 1.3 | Mass and strain fluctuations from Hf/Zr/Ti and nanostructure enhance phonon scattering | (Chen, 2016) |
| $Ti_{0.5}Zr_{0.25}Hf_{0.25}NiSn_{0.98}Sb_{0.02}$ n-type | 1.2 | Phase separation and mass and strain fluctuations | (Gürth, 2016) |
| $Ti_{0.5}Zr_{0.25}Hf_{0.25}NiSn$ n-type | 1.5 | Phase separation and mass and strain fluctuations | (Rogl, 2018) |
| $FeNb_{0.86}Hf_{0.14}Sb$ single dopant p-type | 1.5 | Heavy hole band in FeNbSb enhances thermopower and facilitates high dopant content, Nb/Hf mass fluctuations enhance phonon scattering | (Fu, 2015) |
| $ZrCoBi_{0.65}Sb_{0.15}Sn_{0.20}$ p-type | 1.4 | High hole band degeneracy in ZrCoBi enhances thermopower, Bi/Sb/Sn mass fluctuations and low energy phonons enhance phonon scattering | (Zhu, 2018) |
| $Ta_{0.74}V_{0.1}Ti_{0.16}FeSb$ p-type | 1.52 | High hole band degeneracy in TaFeSb, mass and strain fluctuations, low energy phonon | (Zhu, 2019) |

(B) Entropy-Engineered Chalcogenide Compounds

In this subsection, we shall briefly survey recent chalcogenide thermoelectric materials that are specifically developed using entropy engineering, as summarized in Figure 6 and Table 3. Consistent with the discussion above, these listed MPEA exhibit low lattice thermal conductivity and also low carrier mobility. The success of entropy engineering is hinged upon whether and to what extent the degraded carrier mobility can be compensated, say, by other microscopic electron band parameters such as carrier concentration, band convergence, and band effective mass to name a few.

In one of the earliest reports of high-entropy thermoelectric materials, Shafeie et al.(2015) studied $Al_xCoCrFeNi$ (0.0 <x < 3.0) alloys. They demonstrated the importance of complex microstructures for reducing lattice thermal conductivity and tunable valence electron concentration for controlling electrical transport, despite low ZT values. $Bi_2Te_3$-based materials have been the bench mark thermoelectric material for power generation and solid-state cooling near room temperature. Fan et al. (2016) reported thermoelectric study of a series of $(BiSbTe_{1.5}Se_{1.5})_{1-x}Ag_x$ (x=0-1.2 et%) alloys, in which close-to-theoretical-minimum lattice thermal conductivity ~ 0.47 W/m-K was attained at 400 K and attributed to severe lattice distortion. Doping with a trace amount of Ag effectively enhanced the Seebeck coefficient and further reduced the lattice thermal conductivity. Notably, all samples exhibited surprisingly weak temperature dependence of Hall mobility (on the order of 10-100 $cm^2$/V-sec), not for any known scattering mechanism, which the authors attributed to the effect of severe lattice distortion. At x = 0.9 at%, a peak ZT value of 0.63 was attained at 450 K. The same authors conducted another thermoelectric study in PbSnTeSe alloys (Fan et al., 2017). Again, low thermal conductivity was attained and attributed to strong phonon scattering due to severe lattice distortion. La doping not only enhanced both Seebeck coefficient and electrical conductivity at elevated temperatures, but also suppressed the detrimental bipolar effect. Notably, La doping drastically reduced the Hall mobility from ~ 160 $cm^2$/V-sec in undoped base alloy to 50-80 $cm^2$/V-sec in doped ones. At a La doping ratio of 1.5 at%, a peak ZT value ~ 0.8 was obtained at 873 K. It is important to note that the low lattice thermal conductivity of MPEA may not be fully accounted for by phonon scattering, and the mass and strain field fluctuations may cause phonon broadening, which directly affects the thermal transport (Körmann et al., 2017). In the meantime, the vibrational entropy plays a role as important as the configurational entropy in phase stability (Körmann et al., 2017).

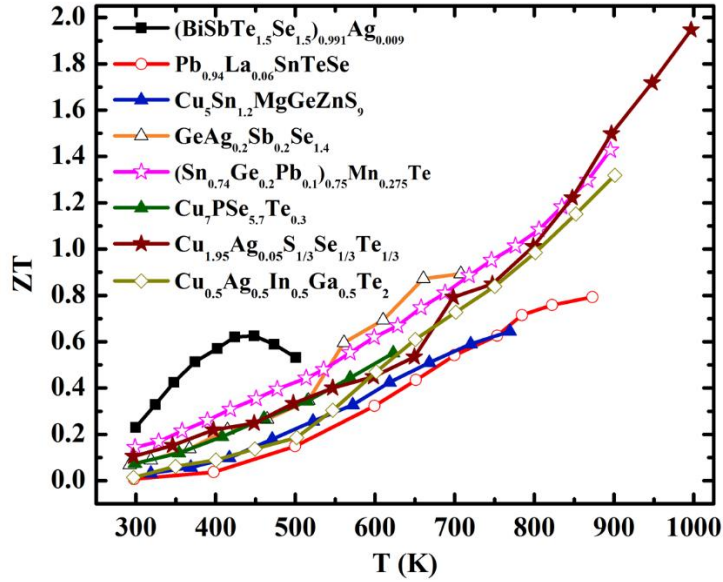

Figure 6. The figures of merit of several chalcogenide MPEA thermoelectric materials as a function of temperature. The references are listed in Table 3.

Table 3. The thermoelectric properties of several chalcogenide MPEA thermoelectric materials at their maximum ZT temperature as indicated. The mobility was derived at room temperature (RT) unless otherwise noted.

| Composition | $\kappa_L$ (W/m-K) | $\sigma$ ($\times 10^4$ S/m) | $\alpha$ ($\mu$V/K) | RT $\mu_H$ (cm$^2$/V-s) | ZT | Enabling mechanism | Refs. |
|---|---|---|---|---|---|---|---|
| $(BiSbTe_{1.5}Se_{1.5})_{0.991}Ag_{0.009}$ p-type (450K) | 0.29 | 1.2 | 220 | 14 | 0.63 | Severe lattice distortion Ag-doping optimized $\alpha$ and $\kappa_L$. | (Fan 2016) |
| $Pb_{0.94}La_{0.06}SnTeSe$ p-type (873 K) | 0.8 | 4.0 | 180 | 65 | 0.8 | Severe lattice distortion La-doping increased $\alpha$, $\sigma$, and suppressed the bipolar effect. | (Fan 2017) |
| $Cu_5Sn_{1.2}MgGeZnS_9$ p-type (773 K) | 0.4 | 5.0 | 128 | | 0.58 | Sn-doping increased $\alpha$ and PF | (Zhang 2018) |
| $GeAg_{0.2}Sb_{0.2}Se_{1.4}$ p-type (710 K) | 0.85 | 1.6 | 265 | 5 | 0.86 | Alloying led to higher lattice symmetry, carrier density, and a multi-valley Fermi surface topology | (Huang 2017) |
| $(Sn_{0.74}Ge_{0.2}Pb_{0.1})_{0.75}Mn_{0.275}Te$ p-type (900 K) | 1.15 | 5.0 | 185 | 15 | 1.42 | Extended solubility of Mn led to enhanced band convergence, effective mass and carrier density. Rich microstructures. Severe | (Hu 2018) |

| | | | | | | lattice distortion | |
|---|---|---|---|---|---|---|---|
| $Cu_7PSe_{5.7}Te_{0.3}$ p-type (600 K) | 0.31 | 0.14 | 238 | 1 | 0.55 | Te doping led to higher lattice symmetry and enhanced α. | (Chen 2018) |
| $Cu_{0.5}Ag_{0.5}In_{0.5}Ga_{0.5}Te_2$ p-type (900 K) | | 0.82 | 343 | | 1.33 | High α, phonon scattering by mass and strain fluctuations | (Liu, 2017) |
| $Cu_{1.95}Ag_{0.05}S_{1/3}Se_{1/3}Te_{1/3}$ p-type (1000 K) | | 2.1 | 240 | | 1.92 | High initial lattice symmetry, native cation disorder, severe lattice distortion | (Liu, 2017) |

Aside from reducing the lattice thermal conductivity, entropy-enabled core effects affect the crystal structure, electron band structure, and the electrical transport as well. For example, doping Te on the Se-site drove the room temperature crystal structure of $Cu_7PSe_6$ from a space group $P2_13$ to a higher-symmetry space group $F\bar{4}3m$ (Chen, 2018). Higher crystal lattice symmetry led to a higher density-of-states effective mass in the same carrier concentration range. As a result, the power factor was drastically increased by a factor of 15, which testifies the effectiveness of entropy-engineering in thermoelectrics. Furthermore, the configurational entropy enabled alloying between compounds with different crystal structures. The room temperature crystal structure of $AgSbSe_2$ and GeSe are cubic rock salt and orthorhombic, respectively. Huang et al. (2017) alloyed $AgSbSe_2$ with GeSe and created new rhombohedral structured GeSe-based alloys that own high band degeneracy and a multi-pocket Fermi surface topology, reminiscent of rhombohedral GeTe. Interestingly, alloying GeSe with Ag or Sb alone didn't stabilize this rhombohedral phase. The alloying-induced new Fermi surface topology and Ge vacancies led to a desired high carrier concentration, close to the optimal one, and a promising ZT ~ 0.86 in $GeAg_{0.2}Sb_{0.2}Se_{1.4}$ at 710 K. Through alloying SnTe with Ge, Pb, Mn and Te vacancies, Hu et al. (2018) reported (i) an extended solubility limit of Mn in (Ge, Pb)-doped SnTe compared to that in pristine SnTe; (ii) a systematic crossover in the carrier scattering mechanism with increasing number of alloying elements, i.e., from electron-phonon scattering to alloy scattering; (iii) the degraded carrier mobility was compensated by increased carrier concentration, heavier effective mass, and band convergence towards a good ZT; and (iv) rich multi-scale microstructures including line defects, strain clusters and dislocation arrays along with point defects led to low lattice thermal conductivity. Fine tuning carrier concentration by Sn excess further boost the peak ZT value to 1.42 at 900 K in $(Sn_{0.74}Ge_{0.2}Pb_{0.1})_{0.75}Mn_{0.275}Te$.

Zhang et al. (2018) developed a data-driven model to optimize the design of ecofriendly thermoelectric high-entropy sulfides. The model, based on data mining of International Crystal Structure Database, assists to prescreen suitable elements to reduce the formation enthalpy and thus facilitate the formation of single phased diamond-like structured MPEA. As a result, metallic $Cu_5SnMgGeZnS_9$ (a solid solution of $Cu_5SnS_4$, $Cu_2MgGeS_4$ and ZnS) and semiconducting $Cu_3SnMgInZnS_7$ (a solid solution of $Cu_2MgSnS_4$, $CuInS_2$ and ZnS) were identified, a ZT of 0.58 was obtained in the latter at 773 K by regulating the carrier concentration via Sn excess. Also in line with the diamond-like structured MPEA, entropy engineering was employed in the study of $CuInTe_2$ via alloying Ag and Ga on the Cu and In site, respectively (Liu et al. 2017). On one hand, co-alloying of Ag and Ga significantly increased the electrical conductivity at elevated temperatures and increased the ZT. On the other hand, it should be noted that the optimal ZT was not attained in the composition with largest configurational entropy.

$Cu_2(S,Se,Te)$ materials are known as "phonon-liquid electron-crystal" thermoelectric materials, a step forward from the classic "phonon-glass electron-crystal" paradigm (Zhao et al. 2019). Outstanding ZT values emerge when these materials enter into a mixed electronic ionic conduction state, in which one sublattice constituted by mobile Cu ions interpenetrates the rigid chalcogen sublattice. Such hybrid sublattices are interesting from configurational entropic perspective: one prerequisite for ionic migration is the presence of available (vacant) sites, over which mobile ions are highly delocalized. Since the available sites generally outnumber the mobile ions, the sublattice of mobile ions has native configurational entropy even without doping/alloying. Meanwhile, one can actively tune the configurational entropy of the rigid chalcogen sublattice and also that of the mobile ion sublattice by doping/alloying. Compared to pristine $Cu_2S$, $Cu_2Se$, and $Cu_2Te$, concomitant with increased configurational entropy are increased ZT. At 1000 K, $Cu_{1.95}Au_{0.05}S_{1/3}Se_{1/3}Te_{1/3}$ exhibits a state-of-the-art ZT ~ 1.92, and the ZT values of $Cu_{1.94}S_{1/2}Se_{1/2}$, $CuS_{1/2}Te_{1/2}$ and $CuS_{1/2}Se_{1/2}$ are ~ 2.25, 2.16 and 2.26 (Zhao et al. 2019), respectively. Special microstructures were discovered in these $Cu_2(S,Se,Te)$-based compounds. On the other hand, the configurational entropy by itself is insufficient to be an indicator of ZT.

VII. FUTURE PROSPECTS

Despite at its early stage, the synergy of two major classes of functional materials: high entropy alloys and thermoelectric materials has paved the way towards "high-entropy thermoelectrics". At the core of this emerging paradigm are four configurational entropy-enabled core effects: extended solubility limit and stabilization of higher crystal lattice symmetry; severe lattice distortion; sluggish diffusion kinetics, and the synergistic effect. In this article, we have surveyed a number of thermoelectric materials from historical perspective and also from configurational entropic perspective. On one hand, the efficacy of extended solubility limit (thus a larger phase space for compositional optimization) and stabilization of higher crystal lattice symmetry (thus a thermoelectrically favorable band structure) is demonstrated. On the other hand, it is uneasy to attribute any of these effects on the crystal structure, microstructures, and transport properties solely to the configurational entropy in presence of other types of entropy and non-entropy factors. The results of entropy engineering study indicated that the configurational entropy needs to be high enough to elicit one or more core effects but low enough to retain a reasonable carrier mobility. In other words, the configurational entropy is a means but not the goal. Hence, multiple principal element alloys, a weaker version of high entropy alloys, are justified. Specifically, the degraded carrier mobility in multiple principal element alloys must be compensated by band convergence, effective mass, and/or carrier concentration to attain competitive thermoelectric performance. In the future, it is desired to inspect the impacts of entropic effects on mechanical properties such as yield strength, toughness, and ductility toward enhanced device-level performance.

ACKNOWLEDGMENTS

S.J.P. thanks Prof. Terry Tritt for many insightful discussion throughout their collaboration on the study of high-entropy half-Heusler compounds. J. H. thanks Dr. Aijuan Zhang for her assistance in preparing this manuscript and valuable comments.